-----------------------------------------------------------------------------
\documentstyle[12pt]{article}
\begin{document}

\begin{flushright}
BIHEP-TH-94-35\\
August,199
\end{flushright}

\begin{center}
 {\large\bf	 Light Flavor Dependence of the Isgur-Wise  Function }
\end{center}
\vspace{0.5cm}
\begin{center}
{\bf Tao  Huang  and  Chuan-Wang Luo}\\
{\small\sl Institute of High Energy Physics, P.O.Box 918(4), Beijing 100039,
China}\\
{\small\sl ( presented by T.Huang )}
\end{center}
\vspace{0.6cm}

\begin{abstract}
	We present an investigation on the ligh flavor dependence of
the Isgur-Wise function for
$B_a\rightarrow D_a$ and $B_a\rightarrow D^{*}_a$ in the framework of
QCD sum rules. It is found that the Isgur-Wise function for $B_s$
decay falls faster than that for $B_{u,d}$ decay , which is just contrary
 to the recent prediction of the heavy meson chiral perturbation theory.
SU(3) symmetry breaking effects in the mass and the decay constant are
also estimated.

\end {abstract}

	As a heavy quark goes into the infinite mass limit, all form
factors for $B\rightarrow D$ and
$B\rightarrow D^{*}$ can be expressed in terms of a single universal function
[1], the so-called Isgur-Wise function.	The Isgur-Wise function
represents the nonperturbative dynamics of weak
 decays of heavy mesons. It depends not only on the dimensionless product
 $v\cdot v^{\prime}$ of the initial and final mesonic velocities, but also on
 the light quark flavor of the initial and final mesons [1-4].
Here, we apply QCD sum rule approach [3,4] to study its light quark flavor
dependence.

       In HQET, the low energy parameter $F_a(\mu)$ of heavy meson
$M_a(\bar{q}Q)$ is defined by [3]
\begin{equation}
 <0|\bar{q}\Gamma h_{Q}|M_a(v)>\;=\;\frac{F_a(\mu)}{2} Tr[\Gamma M(v)].
 \end{equation}
In the leading order, the decay constant $f_{M_a}\simeq
F_a(\mu)/\sqrt{m_{M_a}}$.
It should be emphasized that here and after, the subscript $ a=u,d,s$ specifies
the light antiquark  $\bar{q}=\bar{u},\bar{d},\bar{s}$ of the heavy meson
$M_a(\bar{q}Q)$.

Starting from the
two-point correlation function in HQET,
\begin{equation}
    \pi_5(\omega)\;=\;i\int d^4 x e^{i k\cdot x}<0|T{A^{(v)}_5(x),
A^{(v)+}_5(0)}|0>,
\end{equation}
where $A_5^{(v)}=\bar{q}\gamma_{5} h_{Q}$ and $\omega=2 k\cdot v$  ,
 one obtains the sum rule for $F_a(\mu)$
\begin{eqnarray}
  F_a^2(\mu) e^{-2\bar{\Lambda}_a/T} & = &\frac{3}{8\pi^2}\int^{\omega^c_a}_
{2m_q}ds \sqrt{s^2-4m^2_q}[2 m_q + s]e^{-s/T}
 \nonumber  \\
& &  -<0|\bar{q}q|0>[1-\frac{m_q}{2T}+\frac{m^2_q}{2T^2}]
-\frac{<0|\frac{\alpha_s}{\pi}GG|0>m_q}{4T^2}[\gamma-0.5-ln\frac{T}{\mu}]
   \nonumber   \\
& & +\frac{g_s <0|\bar{q}\sigma Gq|0>}{4T^2}+ \frac{4\pi\alpha_s}{81T^3}
<0|\bar{q}q|0>^2
\end{eqnarray}
with $\gamma=0.5772 $. Taking the derivative with respect to the inverse of T,
  one can  obtain the corresponding sum rule for $\bar{\Lambda}_a$.

	The Isgur-Wise function $\xi_a(v\cdot v^{\prime},\mu)$ is defined
 by the matrix element at the leading order in $\frac{1}{m_Q}$[3],
\begin{equation}
<M_a(v^{\prime})|\bar{h}_{Q_2}(v^{\prime})\Gamma h_{Q_1}(v)|M_a(v)>\;=\;
-\xi_a(v\cdot v^{\prime}, \mu) Tr[\bar{M}(v^{\prime})\Gamma M(v)].
\end{equation}
Same as above,it is not difficult to get the sum rule for the Isgur-Wise
function
\begin{equation}
\xi_a(y,\mu)\;=\;\frac{K(T,\omega^c_a,y)}{K(T,\omega^c_a,1)},
\end{equation}
where
\begin{eqnarray}
K(T,\omega^c_a,y) = & \frac{3}{8\pi^2}(\frac{2}{1+y})^2  \int^{\omega^c}_{m_q
\sqrt{2(1+y)}}   d\alpha [\alpha+(1+y)m_q]\sqrt{\alpha^2-2(1+y)m^2_q}
 e^{-\alpha/T}   \nonumber   \\
 &  -<0|\bar{q}q|0>[1-\frac{m_q}{2T}+\frac{m^2_q}{4T^2}(1+y)]  \nonumber  \\
 & + <0|\frac{\alpha_s}{\pi}GG|0>[\frac{y-1}{48T(1+y)}-\frac{m_q}{4T^2}
(\gamma-0.5-\ln\frac{T}{\mu})]  \nonumber   \\
 & + \frac{g_s<0|\bar{q}\sigma Gq|0>}{4T^2}\frac{2y+1}{3}+
\frac{4\pi\alpha_s <0|\bar{q}q|0>^2}{81T^3} y.
\end{eqnarray}
In the above derivation,  we have used the sum rule  for $F_a(\mu)$.

	In the numerical analysis of sum rules,	we take the parameters
such as condensates and $m_q$ as in [3-4]
and set the scale $\mu=1GeV$. For the continuum model $\omega^c=
\sigma (y)\omega^c_a$, we use the experiment preferred model $\sigma(y)=
\frac{y+1}{2y}$ as  in [3].

	Evaluations of sum rules for $F_a$ and $\bar{\Lambda}_a$ give
\begin{eqnarray}
\bar{\Lambda}_s\simeq 0.62\pm 0.07 GeV,& F_s\simeq 0.36\pm 0.05 GeV^{3/2},\\
\bar{\Lambda}_{u,d}\simeq 0.55\pm 0.07 GeV,& F_{u,d}\simeq 0.32
\pm 0.05 GeV^{3/2}.
\end{eqnarray}

However, in order to reduce the errors,
writing the mass difference $\Delta M=m_{M_s}-m_{M_{u,d}}=
\bar{\Lambda}_s-\bar{\Lambda}_{u,d}$
and the ratio $R_F=F_s/F_{u,d}$ with the corresponding sum rules,
one gets $\Delta M=69\pm 5 MeV$,
which is in good agreement with the recent experiment results [5,6]
$m_{B_s}-m_{B}=90 \pm 6 MeV$,$m_{D_s}-m_{D}=99.5 \pm 0.6 MeV$,
and  the ratio $R_F=1.13 \pm 0.01$.

	In Fig.1, the Isgur-Wise function $\xi_s$ is shown as a function of y.
 Obviously,the dependence on the parameters $\omega_s^c$ and T is very weak.
At the center of the sum rule window  T=0.8GeV,  we obtain the slope
parameter $\varrho^2_a$ defined as $\varrho^2_a=-\xi^{\prime}_a(y=1,\mu)$
\begin{equation}
\varrho^2_s\;=\;1.09\pm 0.04,
\end{equation}
the uncertainty is due to the variation of $\omega^c_s$.
One can compare with
\begin{equation}
\varrho^2_{u,d}\;=\;1.01\pm 0.02.
\end{equation}
and find that SU(3) breaking effects in the slope parameter is not
 large but the important thing is
\begin{equation}
\varrho^2_s > \varrho^2_{u,d}.
\end{equation}
This result just indicates that the Isgur-Wise function $\xi_s$ falls faster
than the Isgur-Wise function $\xi_{u,d}$ as shown below.

	In Fig.2, we show $R_{IW}=\xi_s/\xi_{u,d}$ as a function of y at
$T=0.8 GeV$ for different $\omega^c_{u,d}=1.7\sim 2.3 GeV$ and $\omega_s^c
=\omega^c_{u,d}+0.1GeV$. One can find that the ratio $R_{IW}$ displays a soft
dependence on $\omega^c_{u,d,s}$ .At $ y=1.6$ ($q^2=0$ for $B_{u,d}
\rightarrow D_{u,d}+l\nu $), we get from the sum rule
\begin{equation}
R_{IW}\simeq (95\pm 2 )\%,
\end{equation}
where the uncertainty is ascribed to the uncertainty in $\omega^c_{u,d,s}$
and T.

	 In the evaluations of sum rules for $\xi_a$ and $R_{IW}$, the
continuum model is chosen as $\sigma (y)=\frac{y+1}{2y}$.
This may cause large errors in $ \xi_{a}$ and $ R_{IW}$.
As discussed in [3], one knows
\begin{equation}
\sigma_{min}=\frac{y+1-\sqrt{y^2-1}}{2}\leq \sigma (y) \leq \sigma_{max}=1,
\end{equation}
and the model $\sigma_{max}$ and $\sigma_{min}$
respectively constituents the upper bound and the lower bound for $\xi_a$ .
However,for $R_{IW}$,  the model $\sigma_{max}$ and $\sigma_{min}$
just give the lower bound and the upper bound respectively.  Although
different continuum model gives different value for $R_{IW}$, one can find
that all of these values clearly give
\begin{equation}
R_{IW} < 1~~,~~ for~ y\not = 1.
\end{equation}
Therefore we conclude that $R_{IW} < 1 $(for $y\not = 1$) is independent
of the model choice  $\sigma (y)$.

	In summary ,  It is very interesting to find that
the Isgur-Wise function for $ B_s\rightarrow D_s$ falls faster than the
Isgur-Wise function for $B_{u,d}\rightarrow D_{u,d}$, which is just contrary
to the prediction of the heavy meson chiral perturbation theory where only
SU(3) breaking chiral loops are calculated [2]. Our result $R_{IW}\leq 1$
agrees with that of other calculations [7]. It is expected that the future
experiments can test this result and reveal the underlying mechanism of
 SU(3) breaking effects.

\begin{flushleft}
{\large \bf  Figure Captions}
\end{flushleft}

Fig.1: The Isgur-Wise function $\xi_s$ as a function of y . The band
corresponds
       to variations of 	$\omega_s^c$ in $ 1.8GeV\sim 2.4 GeV$ and T in $
       0.7GeV\sim 0.9GeV $.\\

Fig.2: The ratio $R_{IW}=\xi_s/ \xi_{u,d}$ as a function of y at T=0.8GeV
	($\omega^c_s=\omega^c_{u,d}+0.1GeV$): Dashed line:
	$\omega_{u,d}^c=1.7GeV$, Solid line:
	$\omega^c_{u,d}=2.0GeV$, Dotted line: $\omega^c_{u,d}=2.3GeV$. \\

\end{document}